\documentclass[aps,twocolumn,showpacs,superscriptaddress,groupedaddress,nofootinbib]{revtex4} 
\usepackage{graphicx}
\usepackage[sort&compress]{natbib}
\usepackage{subfigure}
\usepackage{amsmath}
\usepackage{amssymb}
\usepackage{amsfonts}
\usepackage{rotating}
\usepackage{cancel}
\usepackage{lipsum}
\usepackage{mathtools}
\usepackage{color}
\usepackage{bbm}
\usepackage{dsfont}
\usepackage{bbold}
\usepackage{multirow}
\usepackage{ulem}
\newcommand{\eps}{\epsilon}
\begin{document}

\title{Correlation function for the $a_0(980)$}
\author{R. Molina}
\email{Raquel.Molina@ific.uv.es}
\affiliation{Departamento de F\'{\i}sica Te\'orica and IFIC,
Centro Mixto Universidad de Valencia-CSIC, Parc Científic UV, C/ Catedrático José Beltrán, 2, 46980 Paterna, Spain}   
\author{Zhi-Wei Liu}
\affiliation{School of Physics, Beihang University, Beijing 102206, China}  
\author{Li-Sheng Geng}
\email{lisheng.geng@buaa.edu.cn}
\affiliation{School of Physics, Beihang University, Beijing 102206, China}
\affiliation{Beijing Key Laboratory of Advanced Nuclear Materials and Physics, Beihang University, Beijing 102206, China}
\affiliation{Peng Huanwu Collaborative Center for Research and Education, Beihang University, Beijing 100191, China}
\affiliation{Southern Center for Nuclear-Science Theory (SCNT), Institute of Modern Physics, Chinese Academy of Sciences, Huizhou 516000, China}
\author{E. Oset}
\email{Eulogio.Oset@ific.uv.es}
\affiliation{Departamento de F\'{\i}sica Te\'orica and IFIC,
Centro Mixto Universidad de Valencia-CSIC, Parc Científic UV, C/ Catedrático José Beltrán, 2, 46980 Paterna, Spain}

\begin{abstract}  
We have conducted a model independent analysis of the $K^+ \bar{K}^0$ pair correlation function obtained from ultra high energy $pp$ collisions, with the aim of extracting the information encoded in it related to the $K\bar{K}$ interaction and the coupled channel $\pi^+ \eta$. With the present large errors at small relative $K^+\bar{K}^0$ momenta, we find that the information obtained about the scattering matrix suffers from large uncertainties. Even then, we are able to show that the data imply the existence of the $a_0$ resonance, $a_0(980)$, showing as a strong cusp close to the $K\bar{K}$ threshold. We also mention that the measurement of the $\pi^+ \eta$ correlation function will be essential in order to constrain more the information on $K\bar{K}$ dynamics that can be obtained from correlation functions. 

\end{abstract}
\maketitle
\section{Introduction}

Femtoscopic correlation functions are emerging as a powerful tool to learn about hadron interactions. In $pp$ or $pA$ collisions at very high energy, pairs of particles are measured and their production probability is divided by the equivalent probability of uncorrelated events, evaluated through the mixed event method \cite{nakada,star,note,wetzels,toia,kaskulov}. There is already much experimental work on this field~\cite{alice1,alice2,alice3,alice4,alice5,alice6,alice7,alice8,alice9,alice10,stara,starb,fab} and theoretical work goes parallel \cite{morita,ohnishi,feijoc,morita1,hatsuda,mihaylov,haiden,morita2,kamiya,kamiya1,kamiya2,liuwei,vidafe,albal,Liulu,liuwen}, showing that significant information about the interaction of the measured pairs is obtained from correlation functions. While most theoretical works compare models with the results of the correlation functions, it has only been recently that model independent methods have been proposed to extract the information encoded in these correlation functions \cite{ikeno,scibull,Feijootbb,Molinanstar}, such as to conclude the possible existence of bound states, and determine scattering parameters like the scattering length and effective range of the involved channels in the interactions. 

In the present work we take advantage of the existing measurements of the correlation function for $K^0_sK^\pm$ production \cite{expe1,expe2} and extract from there properties about the $K^0K^-$, $\bar{K}^0K^+$ interaction, among them, the existence of a resonance very close to the $K\bar{K}$ threshold, the $a_0(980)$. Uncertainties in observables obtained from the correlation function are evaluated through the resampling method, determining the precision demanded on the experimental data in order to get more precise values of these observables. 

There is already some theoretical work on this correlation function using data from $Pb-Pb$ collisions~\cite{Achasovtwokaon} and applying the Lednicky-Lyuboshitz approximation \cite{ll1,ll2} where a good reproduction of the data was obtained by using different conventions for the correlation function, that did not allow the authors to be very conclusive on the information that can be obtained from these data. Here we employ instead an improved theoretical approach based on the original Koonin-Pratt formula \cite{kp1,kp2}, modified to include the range of the interaction \cite{vidafe} and take the data from $pp$ collisions of the more recent paper \cite{expe2}
\section{Formalism}
\subsection{Brief summary of the chiral unitary approach}
In the chiral unitary approach \cite{npa,kaiser, marku, juanenri}, one uses the $K\bar{K}$ and $\pi\eta$ coupled channels and the scattering matrix given by,
\begin{equation}
 T=[1-VG]^{-1}V\ ,\label{eq:bethe}
\end{equation}
with $V$ given by \cite{xiedai},
\begin{align}
 &V_{K^+K^-,\pi^0\eta}=-\frac{\sqrt{3}}{12f^2}(3s-\frac{8}{3}m^2_K-\frac{1}{3}m_\pi^2-m_\eta^2)\ ,\nonumber\\
 &V_{K^0\bar{K}^0,\pi^0\eta}=-V_{K^+K^-,\pi^0\eta}\ ,\nonumber\\
 &V_{\pi^0\eta,\pi^0\eta}=-\frac{m^2_\pi}{3f^2}\ ,\nonumber\\
 &V_{K^+K^-, K^+K^-}=-\frac{s}{2f^2}\ ,\nonumber\\
 &V_{K^+K^-,K^0\bar{K}^0}=-\frac{s}{4f^2}\ ,\nonumber\\
 &V_{K^0\bar{K}^0, K^0\bar{K}^0}=-\frac{s}{2f^2}\ .\nonumber\\
 \label{eq:pot1}
\end{align}
Given the isospin multiplets: $(K^+,K^0)$, $(\bar{K}^0, -K^-)$, $(-\pi^+,\pi^0,\pi^-)$, one can easily find, for the channels, (1) $K^+\bar{K}^0$, and (2) $\pi^+\eta$,
\begin{align}
 &V_{K^+\bar{K}^0,K^+\bar{K}^0}=-\frac{s}{4f^2}\ ,\nonumber\\
 &V_{K^+\bar{K}^0,\pi^0,\eta}=\frac{\sqrt{3}}{12f^2}(3s-\frac{8}{3}m^2_K-\frac{1}{3}m_\pi^2-m_\eta^2)\ ,\nonumber\\
 &V_{\pi^0\eta,\pi^0\eta}=-\frac{m^2_\pi}{3f^2}\ .
 \label{eq:pot2}
\end{align}
In Eq.~(\ref{eq:bethe}), $G$ is the diagonal meson-meson loop function, with the diagonal elements, $G_{i}$, $i=1,2$, given by,
\begin{align}
 \scriptsize{G(s)=\int^{\scriptsize{q_{\mathrm{max}}}}\frac{d^3q}{(2\pi)^3}\frac{\omega_1(q)+\omega_2(q)}{2\omega_1(q)\omega_2(q)}\frac{1}{s-(\omega_1(q)+\omega_2(q))^2+i\eps}}
 \label{eq:loop}
\end{align}
being $q_\mathrm{max}$ is a regulator of the loop function which in \cite{xiedai} is taken around $600-700$~MeV. The $T$-matrix gives rise to a cusp like structure around the $K\bar{K}$ threshold, in agreement with the shapes obtained in recent experiments \cite{rubin,karnicer}. The approach allows to reproduce different experiments where the $a_0(980)$ is produced, as the $\chi_{c1}\to \eta\pi^+\pi^-$ \cite{liangxie}, the $D_s^+\to \pi^+\pi^0\eta$ decay \cite{raquelgeng}, $D^0\to K^-\pi^+\eta$ \cite{ikegeng}, among others.
\subsection{Correlation functions}
Following Ref.~\cite{scibull} we write the theoretical correlation function,
\begin{align}
 &C_{\bar{K}^0K^+}(p_{K^+})=1+4\pi\theta(q_\mathrm{max}-p_{K^+})\nonumber\\&\times \int^\infty_{0}dr r^2 S_{12}(r)\left\{\vert j_0(p_{K^+}r)+T_{11}(E)\tilde{G}_{1}(r;E)\vert^2\right.\nonumber\\&+\left.\vert T_{21}(E)\tilde{G}_{2}(r,E)\vert^2-j_0^2(p_{K^+}r)\right\}
 \label{eq:cor}
\end{align}
with 
\begin{equation}
 S_{12}(r)=\frac{1}{\left(R\sqrt{4\pi}\right)^3}e^{-\frac{r^2}{4R^2}}\ ,
 \label{eq:so}
\end{equation}
and,
%\begin{widetext}
\begin{align}
 &\tilde{G}_{i}(r,E)=\int\frac{d^3q}{(2\pi)^3}\frac{\omega_{1}+\omega_{2}}{2\omega_{1}\omega_{2}}\frac{j_0(qr)}{s-(\omega_{1}+\omega_{2})^2+i\eps}\ ,
\end{align}
%\end{widetext}
referring the subindices $1,2$ to the two particles in channel $i$, and $\omega_{l}=\sqrt{q^2+m_l^2}$. In Eq.~(\ref{eq:so}), $R$ is the size of the source function where the particles are assumed to be formed. 
\subsection{Inverse problem}
We will make a fit to the data of the correlation function without assumming any specific interaction. The scattering matrix between the two channels, $K^+\bar{K}^0$, and $\pi^+\eta$, is given by Eq.~(\ref{eq:bethe}), taking a general $V$ function,
\begin{eqnarray}
 V=\left(\begin{array}{cc}
   V_{11}&V_{12}\\V_{21}&V_{22}       
         \end{array}
\right)\ .\label{eq:vg}
\end{eqnarray}
Since we are only interested in the region close to $K\bar{K}$, we  assume that the elements $V_{ij}$ are constants. Then, in order to calculate $T$ and the correlation function we have a total of five parameters, three coefficients $V_{ij}$, $q_{\mathrm{max}}$ and $R$. These parameters are fitted to the data.

When performing a fit to the data, one obtains the $5$ parameters with large errors. This is due to the large errors of the data at small momenta but also to the strong correlations between the parameters\footnote{It is  easy to see by using one channel the existence of strong correlations between the $V_{ij}$ parameters and $q_\mathrm{max}$. Indeed, in that case, $T^{-1}=V^{-1}-G$, and we can change $V$ and $G$ simultaneously (through $q_\mathrm{max}$), such that $V^{-1}-G$ does not change at the $K\bar{K}$ threshold.}. In order to quantify uncertainties in the observables, we conduct a bootstrap procedure~\cite{bootstrap,bootstrap2,Albaladejojido}, generating random centroids normally distributed and with the same error bars. Then we carry a fit in each case, determine the parameters and from them the values of the observables. After that, we calculate their dispersion, which gives us the uncertainties with which we can hope to obtain these magnitudes. 
\subsection{Observables}
As we will show in the next section, we get a cusp-like structure corresponding to the $a_0(980)$ resonance in the $|T_{11}|^2$ element of the scattering amplitude.  Then, we also evaluate the scattering length and effective range, $a, r_0$ for the $K^+\bar{K}^0$ and $\pi^+\eta$ channels, given by~\cite{ikeno}
\begin{eqnarray}
 -\frac{1}{a}&=&-8\pi\sqrt{s}T^{-1}\vert_{s=s_\mathrm{th}}\nonumber\\
 r_0&=&\frac{2}{\mu}\left[\sqrt{s}\frac{\partial}{\partial s}\left(-8\pi\sqrt{s}T^{-1}+ik\right)\right]_{s=s_{\mathrm{th}}}\ ,
\end{eqnarray}
with 
\begin{equation}
 k=\frac{\lambda^{1/2}(s,m_1^2,m_2^2)}{2\sqrt{s}}\ ,\label{eq:corn}
\end{equation}
evaluated for $T_{11}$ and $T_{22}$. Since for $K^+\bar{K}^0$ the channel $\pi^+\eta$ is open for decay, the values of the scattering parameters, $a,r_0$, are complex for this channel, while for the $\pi^+\eta$ channel, they must be real. 
\section{Results}
In Fig. 1 we show the correlation function obtained from Fig.~5 of Ref.~\cite{expe2}, taking an average of the data for $\sqrt{s}_{pp}=5.02$~TeV and $\sqrt{s}_{pp}=13$~TeV, and summing the systematic uncertainties (given by boxes in Fig.~5 of~\cite{expe2}) to the statistical errors\footnote{The statistical errors are evaluated by taking the maximum between the average of the errors for  $\sqrt{s}_{pp}=5.02$~TeV and $\sqrt{s}_{pp}=13$~TeV and the difference of both centroids at these two energies divided by two.}.
Instead of using the theoretical correlation function given by Eq.~(\ref{eq:cor}), an experimental parameter $\lambda$ \cite{expe2}, called correlation strength, assumming the role of an experimental efficiency, is introduced. The function to be fitted to the data is\footnote{We are thankful to Valentina Mantovani and Albert Feijoo for instructing us on these details.}
\begin{eqnarray}
 \tilde{C}(p_{K^+})=N\left[\lambda C(p_{K^+})+(1-\lambda)F\right]\ ,
\end{eqnarray}
 where $C(p_{K^+})$ is the theoretical correlation function of Eq.~(\ref{eq:cor}), with $N$ a normalization factor around $1$, 
 and $F$ a flat factor also around $1$. We take $F=1$. For $N=1$, when $C(p_{K^+})\simeq 1$, $\tilde{C}(p_{K^+})$ is also $1$ which is the case of the correlation data beyond $300$~MeV/c.

We perform the following fits (where we have at least six free parameters, three $V_{ij}$ elements, $q_{\mathrm{max}}$, $R$, $\lambda$) :
\begin{itemize}
 \item[-] Fit I. $N=1$ and $\lambda$ is a free parameter. 
 \item[-] Fit II. $\lambda$ is restricted in the interval $(0,1)$, as it should be for an experimental efficiency, and $N$ is a free parameter. 
 \item[-] Fit III. $\lambda$ is restricted in the interval $(0,1)$ and we take $N=1$. 
\end{itemize}

\begin{figure}
\begin{center}
 \includegraphics[scale=0.62]{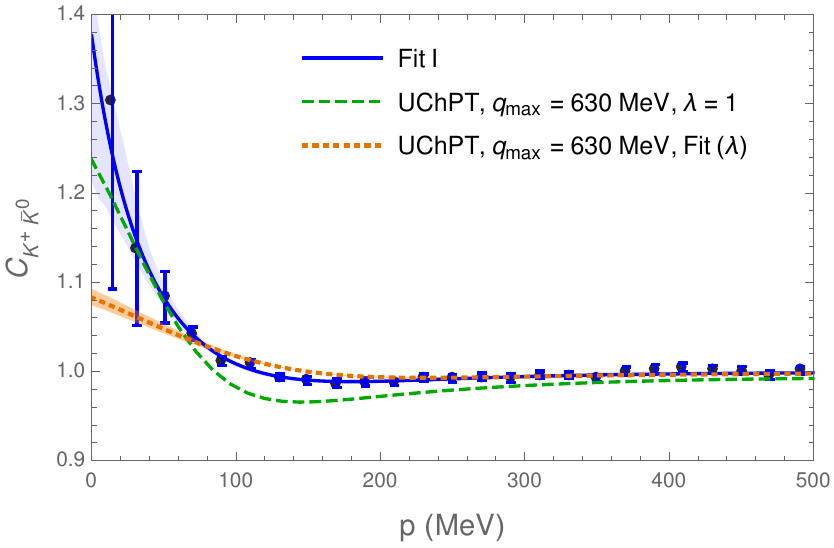}
 \end{center}
 \caption{Comparison of the result of Fit I with the one of UChPT, taking $q_\mathrm{max}=630$~MeV and fixing the $\lambda=1$, or leaving $\lambda$ as a fitting parameter.}
 \label{fig:corfit1}
\end{figure}

\begin{figure}
\begin{center}
 \includegraphics[scale=0.62]{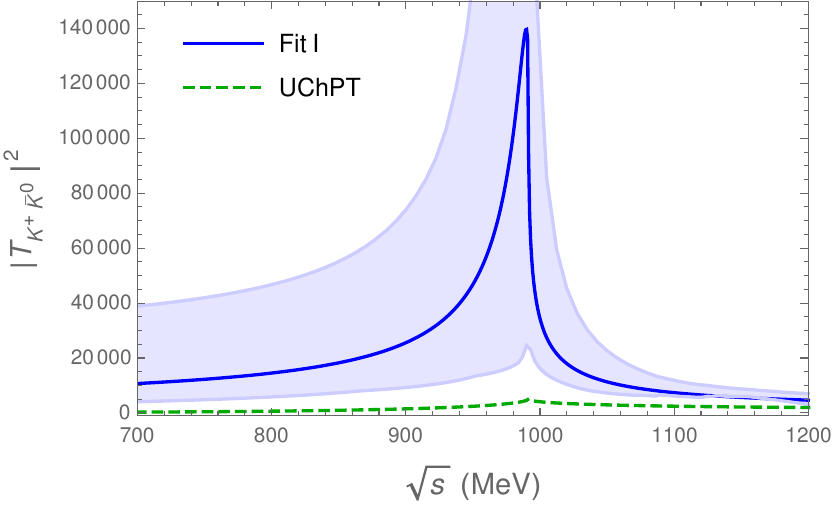}
 \end{center}
 \caption{Comparison of the result of Fit I with the one of UChPT taking $q_\mathrm{max}=630$~MeV.}
 \label{fig:tmatfit1}
\end{figure}

In Fig.~\ref{fig:corfit1} we show the result of Fit I in comparison with the one of UChPT, taking $q_\mathrm{max}=630$~MeV and fixing the $\lambda=1$, or leaving $\lambda$ as a fitting parameter. The value obtained is $\lambda=0.09\pm 0.01$.
We show the $|T_{K^+\bar{K}^0}|^2$ element in Fig.~\ref{fig:tmatfit1}. The best fit is shown as a blue continuous line, while the error band is obtained from the resampling. We obtain a cusp like structure around the $K^+\bar{K}^0$ threshold. However the strength of the peak is more than two orders of magnitude larger than that of the UChPT (which is around $5000$), and once the error band is calculated, it falls much below the result from Fit I. 

In Tables~\ref{tab:para1} and~\ref{tab:para2} we show the values of the parameters of the interaction of the best fit, and also the result for the scattering parameters, $a,r_0$ for the $K^+\bar{K}^0$ and $\pi^+\eta$ channels, with the uncertainties evaluated from the resampling method. Each new fit of the bootstrap procedure returns a set of  parameters that take automatically into account the existing correlations. After every fit we determine the values of the observables, and from many such fits we evaluate the dispersion of these observables.
As we can see, the errors of the free parameters obtained are very large, see Table~\ref{tab:para1}, indicating that there are strong correlations between the parameters. As we have mentioned, this is not a problem since we are interested in the values of the observables, not in those of the parameters.

\begin{figure}
\begin{center}
 \includegraphics[scale=0.62]{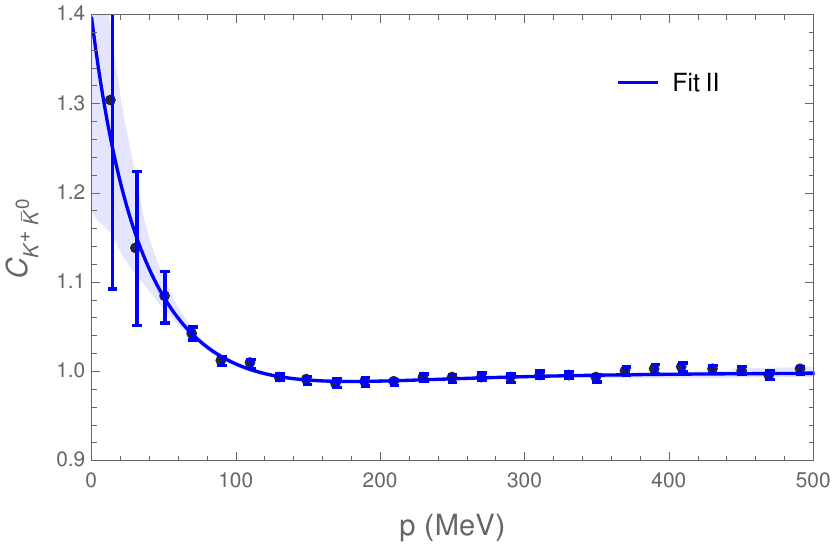}
 \end{center}
 \caption{Result for the correlation function in Fit II.}
 \label{fig:corfit2}
\end{figure}

\begin{figure}
\begin{center}
 \includegraphics[scale=0.62]{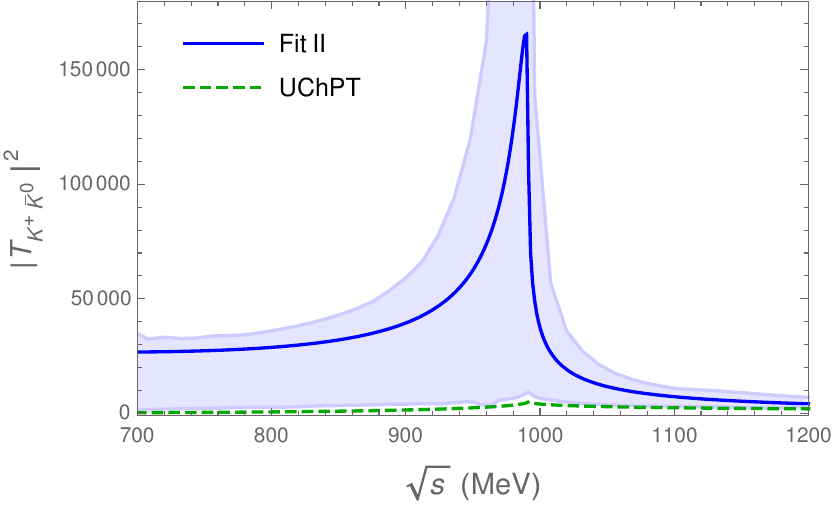}
 \end{center}
 \caption{Comparison of the result of Fit II with the one of UChPT taking $q_\mathrm{max}=630$~MeV.}
 \label{fig:tmatfit2}
\end{figure}

\begin{figure}
\begin{center}
 \includegraphics[scale=0.62]{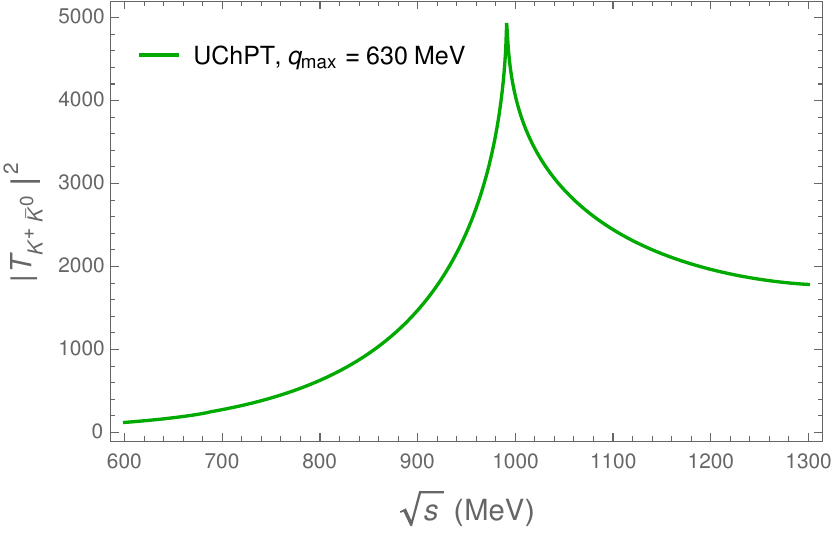}
 \end{center}
 \caption{Absolute squared value of $T_{K^+\bar{K}^0}$ of UChPT taking $q_\mathrm{max}=630$~MeV.}
 \label{fig:tmatchpt}
\end{figure}

We show the results of Fit II also in Tables~\ref{tab:para1} and~\ref{tab:para2}. In this case the $\lambda$ parameter is restricted to the interval $(0,1)$ and the normalization factor $N$ in Eq.~(\ref{eq:corn}) is also a free parameter. The result for the correlation function is shown in Fig.~\ref{fig:corfit2}. The normalization obtained is very close to one, $N=1.000\pm 0.006$, but the error band has become somewhat bigger. The result for the scattering amplitude is shown is in Fig.~\ref{fig:tmatfit2}. Similarly to Fit I, a cusp related to the $a_0(980)$ is clearly visible from the fit. Still the scattering amplitude for UChPT, also shown in Fig.~\ref{fig:tmatchpt} for clarity, is below the result of Fit II, but much closer to the error band than in the case of Fit I. 

Next, we fix the value of $N=1$, and restrict the parameter $\lambda\in(0,1)$, performing Fit III. The results are shown also in Tables~\ref{tab:para1} and~\ref{tab:para2}. These are very similar to the ones of Fit II, but the error bands for the correlation function and scattering amplitude obtained (omitted here since these don't introduce new information) are slightly narrower.

\begin{figure}
\begin{center}
 \includegraphics[scale=0.62]{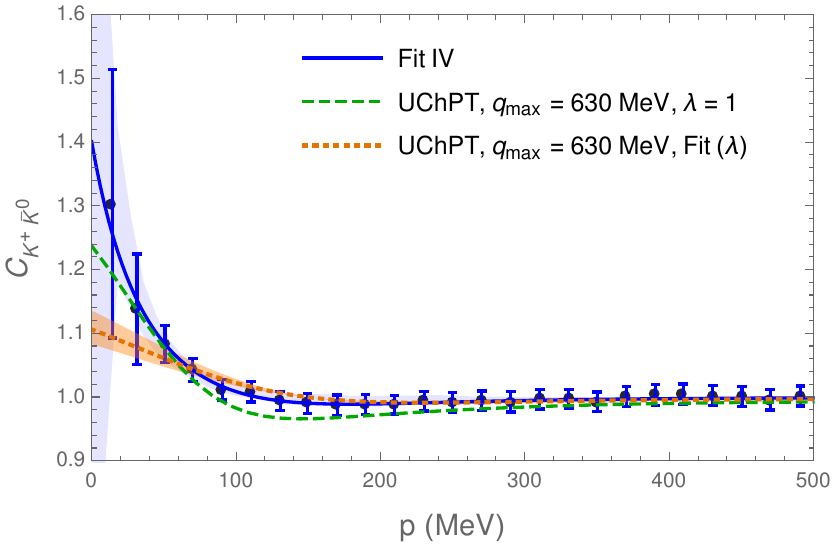}
 \end{center}
 \caption{Result for the correlation function in Fit IV.}
 \label{fig:corfit4}
\end{figure}

\begin{figure}
\begin{center}
 \includegraphics[scale=0.62]{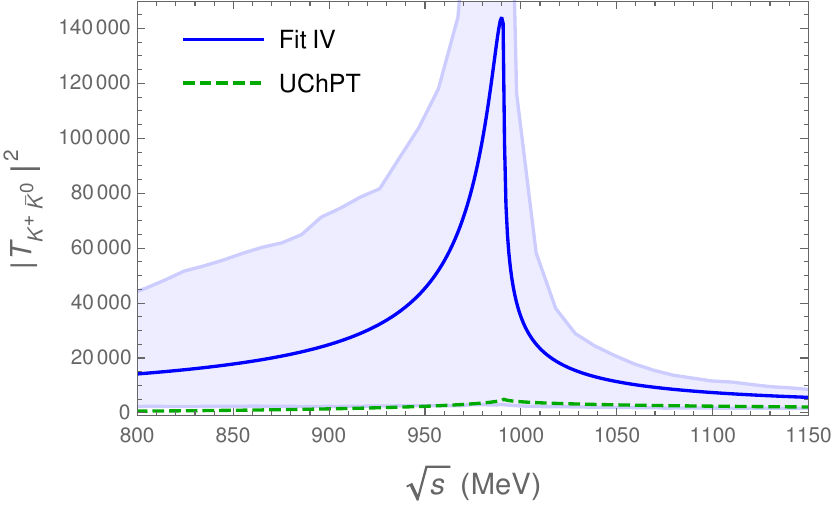}
 \end{center}
 \caption{Comparison of the result of Fit IV with the one of UChPT taking $q_\mathrm{max}=630$~MeV.}
 \label{fig:tmatfit4}
\end{figure}

\begin{figure}
\begin{center}
 \includegraphics[scale=0.62]{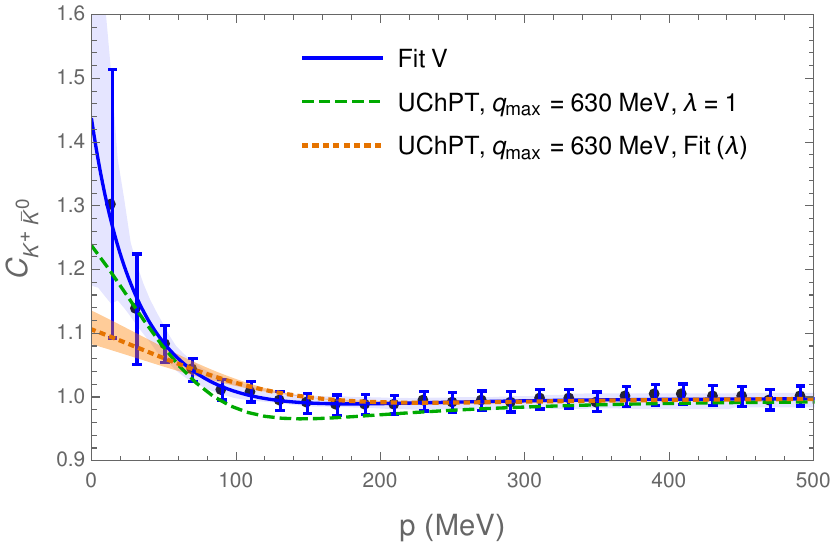}
 \end{center}
 \caption{Result for the correlation function in Fit V.}
 \label{fig:corfit5}
\end{figure}

\begin{figure}
\begin{center}
 \includegraphics[scale=0.62]{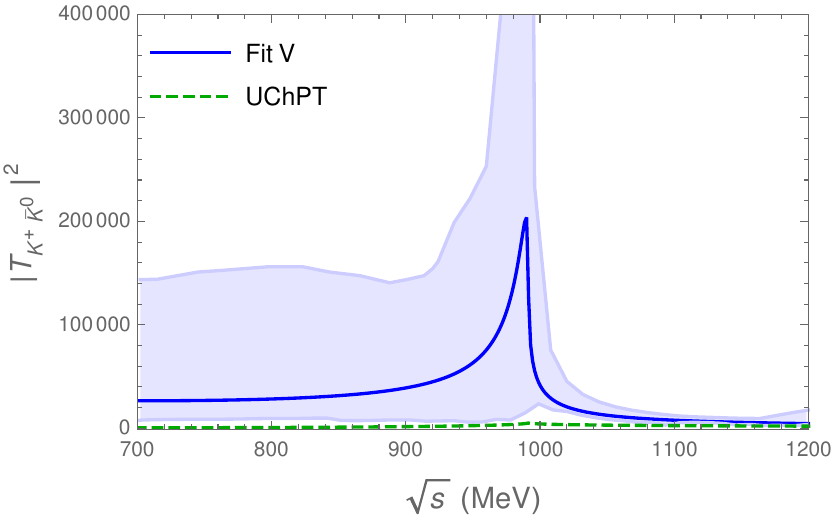}
 \end{center}
 \caption{Comparison of the result of Fit V with the one of UChPT taking $q_\mathrm{max}=630$~MeV.}
 \label{fig:tmatfit5}
\end{figure}

In the following, we make some consideration about the errors of the experimental data. Given that the experimental data of the first three points from Fig.~5 of Ref.~\cite{expe2} contain systematic errors which are indeed quite large, it is surprising that there are not such errors for energies above $p=60$~MeV. It is also shocking that the result of UChPT is completely different in strength, apart from the fact, that both show a cusp close to the mass of the $a_0(980)$. Thus, we propagate the systematic error of the third point around $p=50$~MeV to the rest of data for higher energies, and perform similar fits, labeled as IV, V, VI, in the same way as I, II, III but with the new errors. The results for the free parameters and scattering parameters are shown in Tables~\ref{tab:para1} and~\ref{tab:para2}. The correlation function and scattering amplitude are depicted in Figs.~\ref{fig:corfit4} and \ref{fig:tmatfit4}, in comparison with the UChPT result, where we also show in Fig.~\ref{fig:corfit4} the result from fitting the $\lambda$ parameter for the UChPT case. We obtain $\lambda=0.1\pm 0.03$.
\begin{table}
\renewcommand{\arraystretch}{1.6}
 \setlength{\tabcolsep}{0.03cm}
\begin{center}
\begin{tabular}{ccccccc}
 Fit&$V_{11}$&$V_{22}$&$V_{12}$&$q_\mathrm{max}$&$R$&$\lambda$ \\
 \hline
 I&$-100^{+110}_{-50}$&$-270^{+200}_{-200}$&$-110_{-44}^{+110}$&$1304_{-900}^{+150}$&$0.73^{+0.17}_{-0.08}$&$0.08_{-0.04}^{+0.04}$ \\
 II&$-70^{+80}_{-50}$&$-10^{+340}_{-300}$&$-31^{+30}_{-40}$&$660^{+750}_{-400}$&$0.74^{+0.3}_{-0.06}$&$0.08^{+0.16}_{-0.05}$ \\
 III&$-70^{+60}_{-80}$&$-8^{+310}_{-40}$&$-30^{+180}_{-110}$&$650^{+760}_{-320}$&$0.74^{+0.17}_{-0.07}$&$0.08^{+0.02}_{-0.03}$ \\
 IV&$-100^{+100}_{-100}$&$-270_{-6}^{+520}$&$-110_{-140}^{+140}$&$1272^{+200}_{-800}$&$0.75_{-0.3}^{+9}$&$0.08^{+0.02}_{-20}$ \\
 V&$-71_{-200}^{+16}$&$-13_{-260}^{+120}$&$-29_{-91}^{+110}$&$660_{-390}^{+340}$&$0.80^{+0.8}_{-0.8}$&$0.08^{+0.2}_{-0.1}$ \\
 
 VI&$-72^{+50}_{-10}$&$-12^{+280}_{-280}$&$-29^{+30}_{-140}$&$658^{+400}_{-260}$&$0.75^{+4}_{-0.07}$&$0.08^{+0.04}_{-0.08}$ \\
 \hline
\end{tabular}
 \end{center}
\caption{Values of the parameters obtained from Fits I-VI as described in the manuscript.}
\label{tab:para1}
\end{table}
The error bands obtained in this case are very large, overlapping with the result of fitting the $\lambda$ parameter with UChPT for the correlation function, Fig.~\ref{fig:corfit4}, and also for the scattering amplitude and around the peak, as shown in Fig.~\ref{fig:tmatfit4}. However, as shown in Table~\ref{tab:para1}, in this case we observe that some samples in the resampling lead to value of $\lambda$ out of the range $(0,1)$, that shouldn't be the case in principle\footnote{Note that this only happens when the experimental errors are larger and $\lambda$ is a free parameter as in Fit IV.}. Thus, we conduct Fits V and VI, restricting this parameter to $(0,1)$. In Fit V, we leave $N$ as a free parameter. We obtain, $N=1.0^{+0.01}_{-0.015}$. The results for the correlation function and scattering matrix, shown in Figs.~\ref{fig:corfit5} and~\ref{fig:tmatfit5}, are similar to the previous cases, except for the fact that the error band of the scattering amplitude has become larger for lower energies, and the strength of the peak is slightly higher. 
\begin{table}
\renewcommand{\arraystretch}{1.6}
 \setlength{\tabcolsep}{0.03cm}
\begin{center}
\begin{tabular}{ccccc}
 Fit&$r_1$&$r_2$&$a_1$&$a_2$ \\
 \hline
 I&$0.44_{-0.4}^{-0.4}-i\,0.12_{-0.1}^{+0.1}$&$-0.8^{+0.7}_{-0.7}$&$1.0^{+4.0}_{-1.5}-i\,2.6^{+2.6}_{-0.6}$&$0.45^{+0.4}_{-0.01}$ \\
 II&$0.70^{+0.2}_{-0.3}-i\,0.11^{+0.2}_{-0.2}$&$1.1^{+0.9}_{-0.9}$&$1.2^{+2.7}_{-2.7}-i\,2.7^{+2.7}_{-0.3}$&$-0.3^{+0.9}_{-0.9}$ \\
 III&$0.7^{+0.10}_{-0.14}-i\,0.11_{-0.2}^{-0.2}$&$1.1_{-0.3}^{+0.3}$&$1.3^{+3.4}_{-1.7}-i\,2.7^{+2.3}_{-2.3}$&$-0.25^{+2.0}_{-0.5}$ \\
 IV&$0.45^{+0.3}_{-0.1}-i\,0.11^{+0.14}_{-0.13}$&$-0.73^{+1.3}_{-0.7}$&$1.0^{+2.0}_{-2.5}-i\,2.6^{+2.6}_{-2.6}$&$0.5^{+2.2}_{-0.2}$ \\
 V&$0.71^{+0.2}_{-1.7}-i\,0.09^{+0.1}_{-0.1}$&$1.3^{+1.6}_{-1.6}$&$1.4^{+4.0}_{-6.0}-i\,3.0^{+3.0}_{-3.0}$&$-0.3^{+2.0}_{-2.0}$ \\
 VI&$0.70^{+0.1}_{-0.7}-i\,0.09^{+0.03}_{-0.03}$&$1.3^{+0.8}_{-0.8}$&$1.6^{+1.5}_{-1.5}-i\,2.8^{+1.0}_{-1.0}$&$-0.3^{+1.0}_{-1.0}$ \\
 \hline
\end{tabular}
 \end{center}
\caption{Scattering parameters, $r,a_0$, for Fits I-VI in units of fm for both channels $1$, $K^+\bar{K}^0$ and $2$, $\pi^+\eta$.}
\label{tab:para2}
\end{table}

 \section{Conclusions}
   
 We have used the present data on the correlation function for the $K^+ \bar{K}^0$ $(K-K^0)$ production in high energy $pp$ scattering and have conducted fits to the data with the purpose of extracting the information on the interaction of these pairs encoded in the correlation function. We have used a model independent analysis which provides the scattering amplitudes of these pairs, from where we have seen the structure of these amplitudes and deduced the scattering length and effective range, also for the coupled channel $\pi^+ \eta$. What we observe is that the present accuracy of the data renders an information with very large uncertainties. Within these uncertainties, it is still possible to see that the strength of $\vert T\vert^2$ peaks around the $K\bar{K}$ threshold, indicating that these data corroborates the existence of the $a_0(980)$ resonance, which shows in recent experiments as a strong cusp around the $K\bar{K}$ threshold. It is clear that more precision is needed for the data at small relative momenta of the pair. Yet, this might not be sufficient to get good information. We believe that these data should be complemented with data on the correlation function of $\pi^+ \eta$. We base this conclusion in the experience obtained from the analysis of correlation functions for the $D_{s0}^*(2317)$ \cite{ikeno} and the $N^*(1535)$~\cite{Molinanstar}. Indeed, in the former case, the analysis of the correlation functions of $D^0 K^+$ and $D^+ K^0$ (which are quite similar) allowed one to determine the existence of the $D_{s0}^*(2317)$ state bound by about $40$~MeV, with an accuracy of only $20$~MeV (assuming errors in the data typical of present measurements of the correlation functions). On the other hand, for the latter case, using the correlation functions of the 
$K^0 \Sigma^+$, $K^+ \Sigma^0$, $K^+ \Lambda$ and $\eta p$, one could determine the position of the $N^*(1535)$ state, bound by about $75$~MeV with respect to the $K^+\Lambda$ and $150$~MeV with respect to the $K^+ \Sigma^0$ threshold, with an accuracy of $5$~MeV. It is clear that the combined information of the correlation functions of coupled channels to which the state couples, is far richer than the information from one channel alone. 
    The large uncertainties of the results obtained here are in line with the results of \cite{Achasovtwokaon}, where the correlation function from $Pb-Pb$ collisions was reproduced with multiple, quite different scenarios, to the point that the conclusion of the authors was, 
``as far as we understand, it is not easy to achieve progress in this field''.  Instead of using different models, we have performed a model independent analysis, with a statistical resampling method, that allows to tell us which information we can obtain and with which uncertainties. While these are definitely very large, it is still rewarding to see that the data provide information about the existence of the associated $a_0(980)$ resonance. We are also more positive, in the sense that we show in which direction more information could be obtained, which is, more precision at small relative momenta, and the measurement of the $\pi^+ \eta$ correlation function. 
   One result of the study is that the present data seem to be incompatible, or barely compatible, with the results of chiral unitary theory for the $a_0(980)$ resonance, which has been very successful to explain different reactions where the $a_0(980)$ resonance is explicitly seen. While the results of chiral unitary theory are qualitatively in agreement with the data for the correlation function, the apparent disagreement stems from small discrepancies in the region of $p=100-200$~MeV/c, given the extremely small experimental errors of the data. The success of the chiral unitary approach concerning the $a_0(980)$ should be a reason to revise the data, or most probably the algorithm that should be used to compare with the data. Actually, the procedure to construct the correlation function dividing the probability to find a pair from a single event, by the probability of the mixed events, might require some revision, since the mixed event probability is not fully absent from correlations.

The results and discussion carried here should serve as a motivation to perform new measurements and also look for an adequate theoretical algorithm to compare with experiment. This would also allow one to obtain low energy data for the $K\bar{K}$ and $\pi^+ \eta$ interaction, as scattering lengths and effective ranges, which would be most welcome.

\section{Acknowledgments}

%a careful reading of the paper and 
R. M. acknowledges support from the CIDEGENT program with Ref. CIDEGENT/2019/015, the Spanish Ministerio de Economia
y Competitividad and European Union (NextGenerationEU/PRTR) by the grant with Ref. CNS2022-13614. L. S. G. acknowledges supports from the National
Natural Science Foundation of China under Grants No.11975041 and No.11961141004. This work is also partly supported by the Spanish Ministerio de Economia y Competitividad (MINECO) and
European FEDER funds under Contracts No. FIS2017-84038-C2-1-P B, PID2020-112777GB-I00, and by Generalitat Valenciana under contract PROMETEO/2020/023. This project has received funding from the European Union Horizon 2020 research
and innovation programme under the program H2020-INFRAIA-2018-1, grant agreement No. 824093 of the STRONG-2020
project.

\bibliography{biblio}

\end{document}